%% file: AbsorptionNIR-v09-arxiv.tex
\documentclass[a4paper,review,number,sort&compress,english]{elsarticle}
\usepackage[T1]{fontenc}
\usepackage[latin9]{inputenc}
\usepackage{varioref}
\usepackage{units}
\usepackage{amsmath}
\usepackage{amssymb}
\usepackage{graphicx}
\usepackage{esint}
\usepackage{caption}
\usepackage{subcaption}
\pdfpageheight\paperheight
\pdfpagewidth\paperwidth

\journal{Elsevier}
\usepackage{upgreek}
\usepackage[bookmarks,bookmarksopen,bookmarksdepth=6]{hyperref}
\usepackage{cases}
\usepackage{url}
\usepackage{lineno}
\usepackage{newtxtext,newtxmath,amsmath}
\makeatother

\usepackage{babel}
\begin{document}
%\let\clearpage\relax
%\renewcommand{\thefootnote}{\fnsymbol{footnote}}
% - 3.4.2019 Include comments CS, FF
% 4.4.2019 Version with separate discussion of possible interpretations - plots with alpha_rad/Phi
% Revision 1 - includes changes requested by reviewers

\title{Influence of radiation damage on the absorption of near-infrared light  in silicon}
%\title{}
\author[]{C.~Scharf\corref{cor1}}
\author[]{F.~Feindt}
\author[]{R.~Klanner}

\cortext[cor1]{Corresponding author. Email address: Christian.Scharf@desy.de,
    now at Institut f\"ur Physik, Humboldt-Universit\"at zu Berlin,
 Newtonstr.~15, D~12489 ~Berlin, Germany, Tel. +49~30~2093~7790.}

\address{ Institute for Experimental Physics, University of Hamburg,
 \\Luruper Chaussee 147, D\,22761, Hamburg, Germany.}

%\pdfbookmark[5]{Abstract}{abstract}}

%----- includes corrections of Jack Rohlfs

\begin{abstract}
% \label{sect:Abstract}

 The absorption length, $\lambda _{abs}$, of light with wavelengths between 0.95 and 1.30~$\upmu$m in silicon irradiated with 24~GeV/c protons to 1~MeV neutron equivalent fluences up to $8.6 \times 10^{15}$~cm$^{-2}$ has been measured.
 It is found that $\lambda _{abs}$ decreases with fluence due to radiation-induced defects.
 A phenomenological parametrisation of the radiation-induced change of $\lambda _{abs}$ as a function of wavelength and neutron equivalent fluence at room temperature is given.
 The observation of the decrease of $\lambda _{abs}$ with irradiation is confirmed by edge-TCT measurements on irradiated silicon strip detectors.
 Using the measured wavelength dependence of $\lambda _{abs}$, the change of the silicon band-gap with fluence is determined.

\end{abstract}


\maketitle
 \tableofcontents
%--- \newpage
 \pagenumbering{arabic}

\section{Introduction}
 \label{sect:Introduction}
% Here the introduction

 A standard way of determining the Charge-Collection-Efficiency, $CCE$, of radiation-damaged silicon sensors uses red and near-infrared (NIR) light to generate electron-hole pairs.
 In order to determine the absolute number of produced charge carriers, the light-absorption length, $\lambda _{abs}$, has to be known.
 Radiation produces defect states in the silicon band-gap, which are expected to cause a reduction of $\lambda _{abs}$.

 In Ref.~\cite{Fan:1959} the photo-conductivity of silicon for wavelengths up to $\lambda = 3 ~ \upmu$m has been measured before and after irradiation.
% Whereas before irradiation no photo-conductivity has been observed, it is significant after irradiation.
 Before irradiation no photo-conductivity has been observed.
 After irradiation it becomes significant.
 The results have been described by an irradiation-induced reduction of the silicon band-gap by up to 100~meV.
 However, the fluences of these studies are significantly higher than the ones investigated in this paper, for which a band-gap narrowing of only a few meV is expected.

 In this work the transmittance of high-ohmic $n$-doped silicon irradiated with 24~GeV/c protons up to 1~MeV neutron equivalent fluences, $\Phi _{eq} $, of $8.6~ \times 10^{15}$~cm$^{-2}$ for light with $\lambda $  between 0.95 and 1.30~$\upmu $m has been measured.
 From the results $\lambda _{abs} (\Phi _{eq}, \lambda )$ is derived and the inverse of the radiation-induced absorption length
 \begin{equation}\label{eq:alpha}
   \alpha_{irr} (T, \Phi _{eq}, \lambda ) = 1/\lambda _{abs}(T, \Phi _{eq}, \lambda ) -  1/\lambda _{abs}(T, 0, \lambda)
 \end{equation}
 determined.
 The observed decrease of $\lambda _{abs} $ with $\Phi _{eq} $ is confirmed by  edge-TCT measurement using light from a sub-nanosecond laser with $\lambda = 1.052~\upmu $m~\cite{Scharf:2018,Feindt:2017}.
 The measured dependence of $\lambda _{abs} $ on $\Phi _{eq} $ is also used to investigate a possible narrowing of the silicon energy band-gap, $E_{gap}$, with $\Phi _{eq}$.

 \section{Samples and light-absorption measurements}
  \label{sect:Samples}

 Four samples of phosphorus-doped silicon with $\approx3.5$~k$\Omega $~cm resistivity and $\approx300~\upmu$m thickness were irradiated with 24~GeV/c protons to $\Phi_{eq} = (2.4, 4.9, 6.4, 8.6) \times 10^{15}$~cm$^{-2}$ at the CERN PS.
 The uncertainty of the fluence is about $\pm 10$~\%.
 For the calculation of $\Phi _{eq}$ a hardness factor $\kappa = 0.62$ is  used~\cite{Moll:2002}.
 After irradiation, the samples were stored in a freezer, being only warmed up to room temperature for the measurements; these were performed at $20 \pm 2~^\circ$C.
 A fifth sample, which was not irradiated, provided the results for non-irradiated silicon.
% The crystals were intended for oxide-growth studies.
 The crystals were bare and the thickness of the naturally grown SiO$_2$ layer is much smaller than the wavelength of the light used for the transmission measurements.
% without a SiO$_2$ layer, which would have changed the surface transmittance.
 The mechanical thicknesses of the silicon pieces were measured using a caliper to an accuracy of $2~\upmu$m.

\begin{figure}[!ht]
   \centering
    \includegraphics[width=0.8\textwidth]{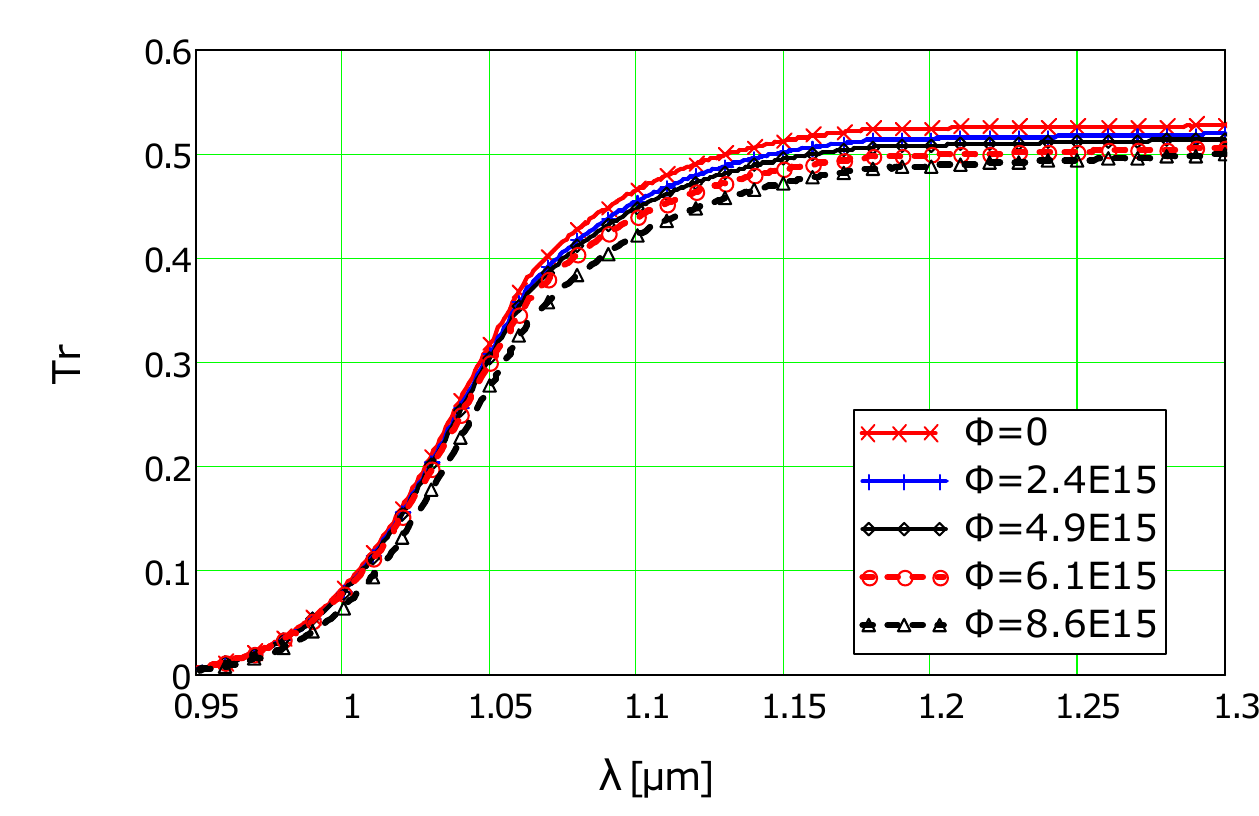}
   \caption{Measured transmittance as a function of wavelength, $\lambda $, for different 1 MeV equivalent neutron fluences, $\Phi _{eq}$, in units of cm$^{-2}$. }
  \label{fig:Transmission}
 \end{figure}

 The transmittance measurements were performed using an Agilent CARY 5000 UV-VIS-NIR spectro-photometer~\cite{Agilent}.
 Fig.~\ref{fig:Transmission} shows the results.
 The transmittance, \emph{Tr}, is close to zero at $\lambda = 0.95~\upmu$m and increases to  $\approx 50~\%$ at $\lambda = 1.30~\upmu$m.
 \emph{Tr}  decreases smoothly with $\Phi _{eq}$.
 From repeated transmittance measurements of the non-irradiated samples a reproducibility of the results at the 0.1~\% level for $\lambda \geq 1.05 ~ \upmu$m is deduced.
 For $\lambda < 1.05~\upmu$m, \emph{Tr} has a strong temperature dependence and the reproducibility worsens to $\approx 1~\%$.

 Next, the transmittance measurements are compared to the expectations from literature values.
 Refs.~\cite{Green:1995,Green:2008} give tables of $n(\lambda )$ and $1/\lambda _{abs} (\lambda) $ and theit temperature dependencies at 300\,K.
 The transmittance, \emph{Tr}$(\lambda )$, of an optical material of thickness $d$, index of refraction $n(\lambda )$ and absorption length $\lambda _{abs} (\lambda )$ is given by
 \begin{equation}\label{equ:Transmission}
   Tr(n, \lambda _{abs} ,d)=\frac{Tra^2 \cdot e^{- d/\lambda _{abs}}} {1 - \Big(Ref \cdot e^{-d/\lambda _{abs}} \Big)^2 },
 \end{equation}
 with the relations for the transmittance, \emph{Tra}$(\lambda)$, and reflectivity, \emph{Ref}$(\lambda) $, of a single air-silicon interface
 \begin{equation}\label{equ:Fresnel}
   Ref(\lambda) =\frac{\big(n(\lambda )-1 \big)^2} { \big(n(\lambda )+1 \big)^2} \hspace{5mm} \mathrm{and} \hspace{5mm}
   Tra(\lambda ) = 1 - Ref(\lambda) =\frac{4 \cdot n(\lambda )} {\big(n(\lambda )+1 \big)^2}.
 \end{equation}
 These are the Fresnel formulae for normal light incidence between a medium of refractive index 1 (air) and $n$ (silicon).
 To derive Eq.~\ref{equ:Transmission}, multiple transmissions and reflections at both air-silicon interfaces have to be considered.
 This results in a geometrical series which, in the limit an infinite number of reflections, gives Eq.~\ref{equ:Transmission}.

 \begin{figure}[!ht]
   \centering
    \includegraphics[width=0.8\textwidth]{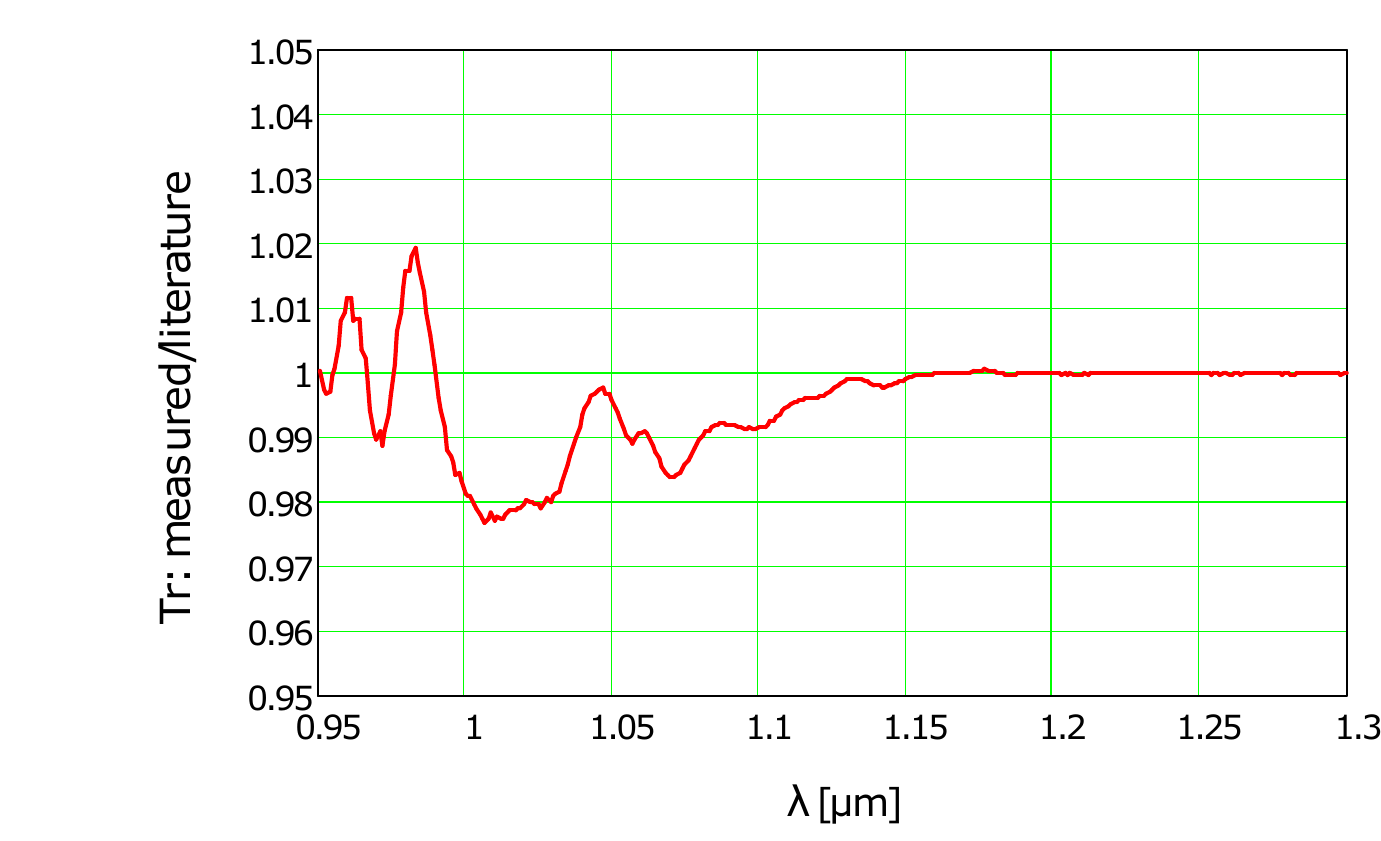}
   \caption{ Ratio of the measured transmittance, $Tr (\lambda )$, to the literature values at $+ 20 ^\circ $C for the non-irradiated silicon.  }
  \label{fig:TrRatio}
 \end{figure}

 Fig.~\ref{fig:TrRatio} shows for the non-irradiated silicon the ratio of the measured to the expected transmittance at $20~^\circ$C, after the following corrections:
% Above  $\lambda = 1.15~\upmu$m the agreement is within 0.1~\%.
% Below $\lambda = 1.15~\upmu$m, \emph{Tr} changes rapidly, and the data are sensitive to the calibration of the $\lambda $-scale and to the temperature.
% In order to get the best agreement, for the figure, the $\lambda $-scale is  shifted by $ 1.0 $~nm and the transmittance scaled by 0.4~\%, which is within the calibration uncertainty of the photospectrometer.
 The $\lambda $-scale is shifted by $ -1.0 $~nm and the transmittance scaled by 0.4~\%, which is within the calibration uncertainty of the photospectrometer.
 With these corrections, the maximum difference is below $\pm 3~\%$.
 Without them, the difference is $+ 15~\%$  at $\lambda = 0.95~\upmu$m, illustrating the high sensitivity of the results to the $\lambda $-scale at short wavelengths.

 \section{Results}
  \label{sect:Results}

 From Eq.~\ref{equ:Transmission} the dependence of $\lambda _{abs}$ on \emph{Tr} can be derived:
 \begin{equation}\label{equ:Labs}
   \lambda _{abs}(\lambda ) = \frac{d} {\ln \Big( \frac{ Tra(\lambda)^2 + \sqrt{Tra(\lambda)^4 + 4 \cdot Ref(\lambda)^2 \cdot Tra(\lambda)^2 }} {2 \cdot Tr(\lambda) } \Big)}.
 \end{equation}

 \begin{figure}[!ht]
   \centering
    \includegraphics[width=0.8\textwidth]{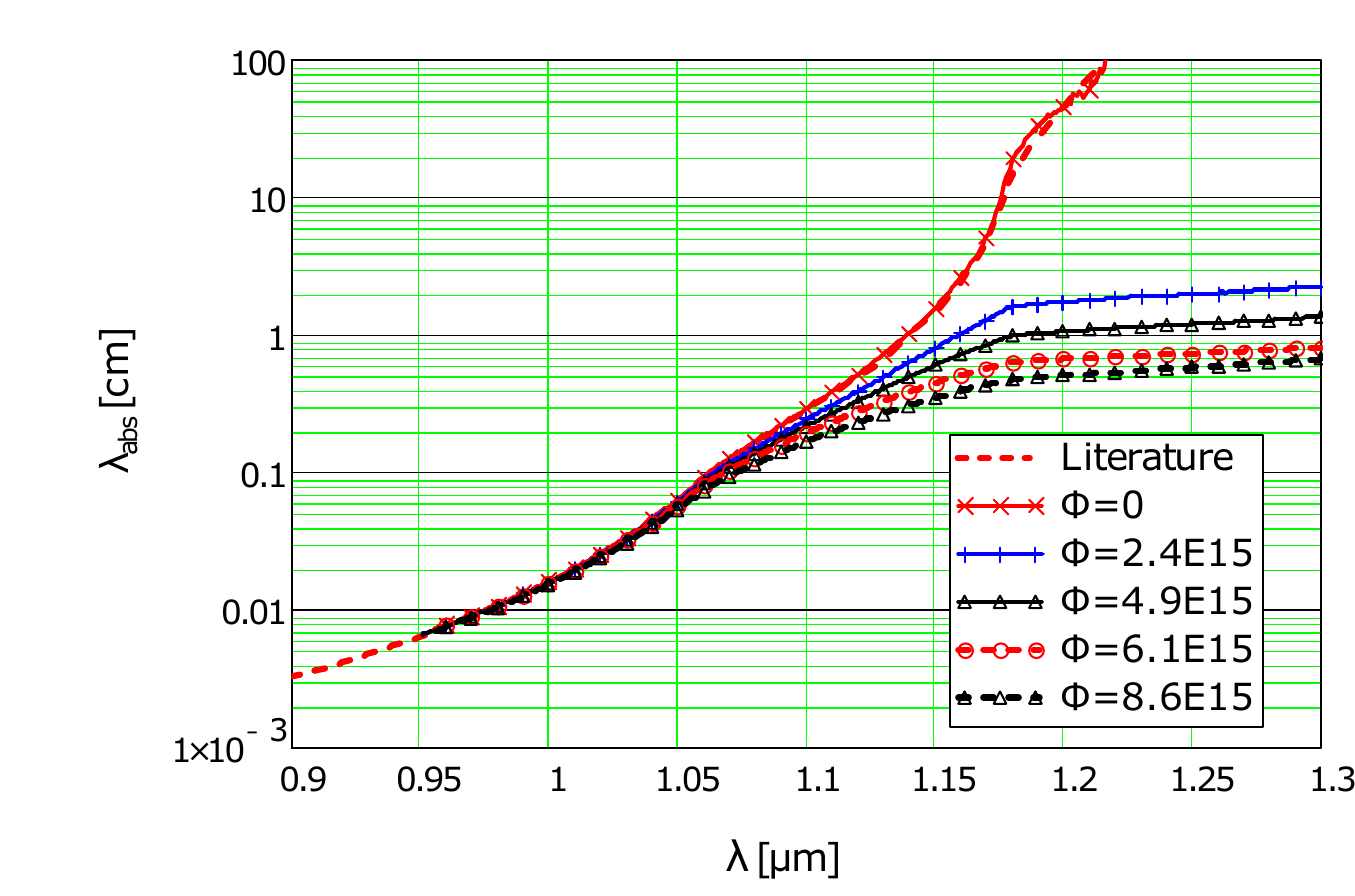}
   \caption{ Measured $\lambda _{abs}$ as a function of $\lambda $ for the different $\Phi _{eq}$-values, and comparison to the values from Ref.\cite{Green:1995} for non-irradiated silicon. }
  \label{fig:Labs}
 \end{figure}

 Fig.~\ref{fig:Labs} shows the experimental results for $\lambda _{abs } (\lambda)$ for the different $\Phi _{eq}$-values, using $n(\lambda )$ at $20~^\circ $C from Ref.~\cite{Green:1995} for both irradiated and non-irradiated silicon.
 In addition, the values for $\lambda _{abs }$ at $20~^\circ $C from Ref.~\cite{Green:1995} are shown.
 The results for the non-irradiated sensor agree with the literature values up to $\lambda = 1.25~\upmu$m.
 For higher values they deviate.
 Note that at $\lambda = 1.2~\upmu$m, $\lambda _{abs } \approx 50$~cm, which is much larger than the sample thickness of 0.03~cm, and the determination of $\lambda _{abs }$ becomes unreliable.
 For the irradiated silicon $\lambda _{abs }$ decreases with $\Phi _{eq}$.

  \begin{figure}[!ht]
   \centering
    \includegraphics[width=0.8\textwidth]{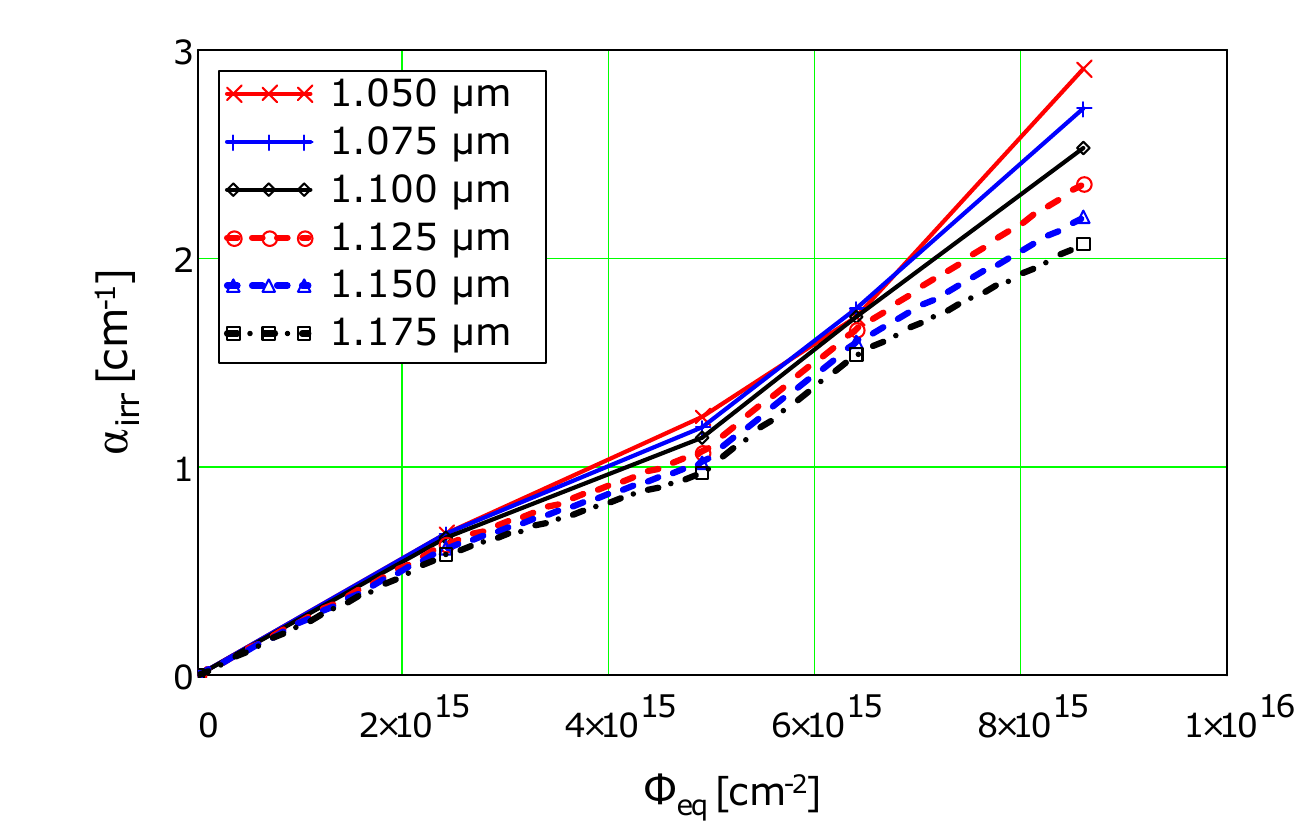}
   \caption{$\alpha _{irr}$ as a function of $\Phi _{eq}$ for selected wavelengths. Within the experimental uncertainties $\alpha _{irr} (\Phi _{eq}) $ is proportional to $\Phi _{eq}$. }
  \label{fig:airr}
 \end{figure}

 The inverse of the radiation-induced absorption length, $\alpha _{irr}$, has been introduced in Eq.~\ref{eq:alpha}.
 Fig.~\ref{fig:airr} shows the experimental results for $\alpha _{irr}(\Phi)$  for selected values of $\lambda $.
% Given the uncertainties of the transmission measurements discussed above, only results for $\lambda \geq 1.05~\upmu$m are shown.
 It is observed that $\alpha _{irr} (\Phi _{eq})$ is approximately proportional to  $\Phi _{eq}$.
% with a slope slowly decreasing with $\lambda $.
 Fig.~\ref{fig:Labs} shows that $\lambda _{abs}$ rapidly decreases with decreasing $\lambda $, and that for small $\lambda$ values, $ \alpha _{irr} $ is the difference of two large numbers.
 As a result, the determination of $\alpha _{irr} $ is very sensitive to the $\lambda $ calibration of the spectrophotometer for $\lambda \lesssim 1.05~\upmu$m.
 Given these uncertainties, only results for $\lambda \geq 1.05~\upmu$m are shown in the following.

\begin{figure}[!ht]
   \centering
    \includegraphics[width=0.8\textwidth]{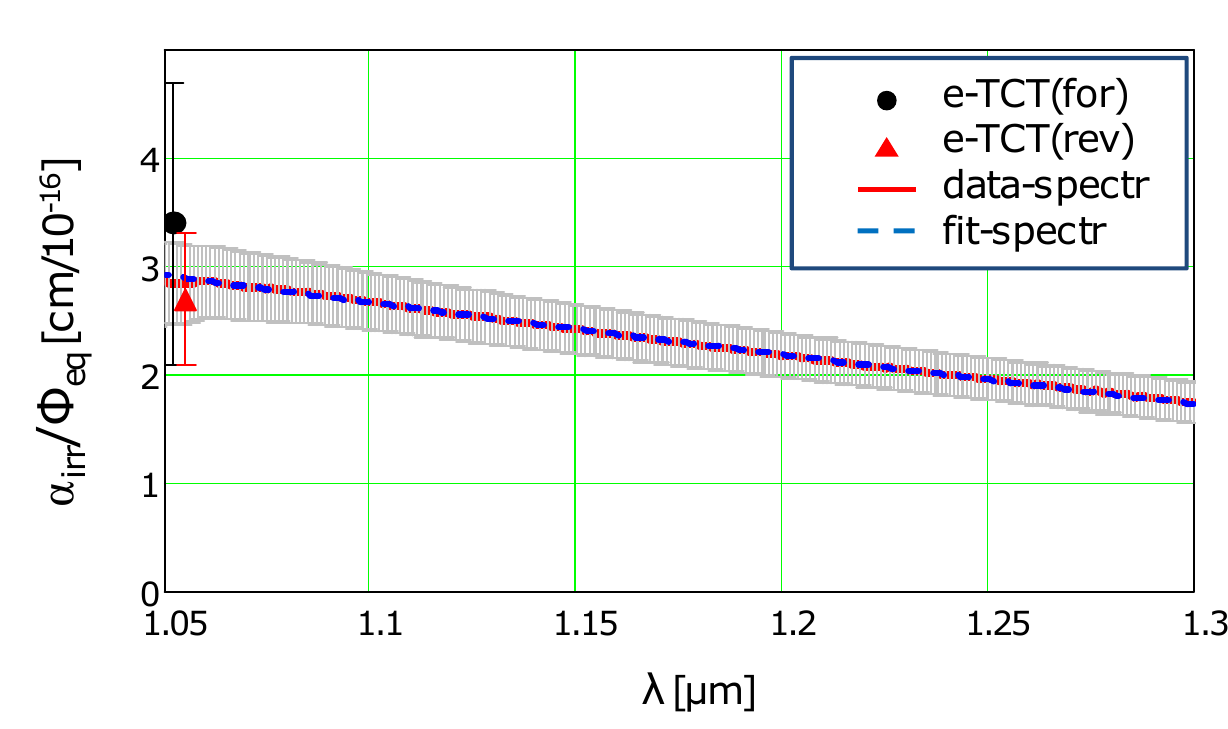}
   \caption{ Results for $\alpha _{irr} / \Phi _{eq}$:
   The edge-TCT results for $\lambda = 1.052~\upmu$m and $-30~^\circ $C are denoted \emph{e-TCT (for)} and \emph{e-TCT (rev)} for forward and reverse voltages, respectively.
   The photospectrometer data at $+20~^\circ $C are labeled \emph{data-spectr}, and the second-order polynomial fit \emph{fit-spectr}.
   For clarity the \emph{e-TCT} results are drawn at different $\lambda $~values.  }
  \label{fig:SigmaBeta}
 \end{figure}

%---move beta x sigma discussion to next section
% $\alpha _{irr} (\lambda )$, which describes the radiation-induced light absorption according to $e^{-\alpha _{irr} (\lambda ) \cdot x}$, may be interpreted as the product of the density of scattering centers, $N_t (\Phi _{eq}, \lambda)$, which can change their charge states by the interaction of a photon of wavelength $\lambda $, and  the photo-absorption cross-section, $\sigma _\gamma $.
% The proportionality $\alpha _{irr} \propto \Phi _{eq}$ suggests the parametrisation $N_t (\Phi _{eq}, \lambda) = \beta _{irr} (\lambda ) \cdot \Phi _{eq}$, with the parameter $\beta _{irr}(\lambda) $ describing the introduction rate of the damage centers.
% Thus the measurements determine
% \begin{equation}\label{equ:IntSig}
%   \big( \sigma _\gamma \cdot \beta _{irr} \big) = \frac{\alpha _{irr} } {\Phi _{eq}},
% \end{equation}
% which allows calculating $\alpha _{irr} $ for a given $\lambda $ and $\Phi _{eq}$.
% Fig.~\ref{fig:SigmaBeta} shows $\sigma _\gamma \cdot \beta _{irr}$ for $\lambda = 1.05$ to $1.30~\upmu$m.
 The proportionality constants, $\alpha _{irr}/\Phi_ {eq}$,  are obtained by calculating the mean $\alpha_{irr}(\lambda, \Phi_{eq})/\Phi_{eq}$  for every $\lambda $-value.
 Fig.~\ref{fig:SigmaBeta} shows the results for $\lambda = 1.05$ to $1.30~\upmu$m.
 The error bars reflect the uncertainty obtained from the straight-line fits to $\alpha _{irr} (\Phi _{eq})$.
 Other sources of uncertainties considered are:  1~\% for $n$, 2~\% for \emph{Tr}, 2~$\upmu $m for the sensor thickness, and 10~\% for the uncertainty of $\Phi _{eq}$.
 The latter one, which is the same for all wavelengths, dominates.
% Additional systematic errors from  uncertainties of 1~\% for $n$, 2~\% for \emph{Tr} and 2~$\upmu $m for the sensor thickness, are 1~\% or less, and 10~\% for the uncertainty of $\Phi _{eq}$.
 The results of a fit by a second order polynomial to the data are given in Table~\ref{tab:BetaSigma}.
 The parameters can be used to calculate $\alpha _{irr} (\Phi _{eq}, \lambda )$,
 and together with Eq.~\ref{eq:alpha} and the values of $\lambda _{abs} (0, \lambda)$ from Ref.~\cite{Green:2008}, the values of $\lambda _{abs} (\Phi _{eq}, \lambda)$.
 With the assumption that the number of electron-hole pairs is proportional to the light absorption, the number of charge carriers generated by light in radiation-damaged silicon at 20~$^\circ $C can be obtained.

 \begin{table} [!ht]
  \centering

   \begin{tabular}{c|c|c}
%   \hline
     $a$ [cm] & $b$ [cm$/\upmu$m]  & $c$ [cm$/\upmu$m$^2$] \\
     \hline
     $(2.427 \pm 0.001 \pm 0.4 \pm 0.2) \times 10^{-16}$ & $(4.805 \pm 0.005) \times 10^{-16}$ & $(1.29 \pm 0.02)\times 10^{-16}$ \\
%   \hline
   \end{tabular}
    \caption{Results of the second-order polynomial fit
      $\alpha _{irr} / \Phi _{eq} = a + b \cdot (\lambda - \lambda_0) +  c \cdot (\lambda - \lambda_0)^2 $
      to the data shown in Fig.~\ref{fig:SigmaBeta}, with the value $\lambda _0 = 1.15~\upmu$m.
      The first error of $a$ is the statistical uncertainty, the second  an estimate of the systematic uncertainty of the analysis method, and the third the result of other systematic effects, which are discussed in the text.
    \label{tab:BetaSigma}}
  \end{table}

%% Typical radiation-induced introduction rates determined with microscopic techniques ~\cite{Davies:2006, Moll:2002} are 1 to 10~cm$^{-1}$.
% The observed $\sigma _\gamma \cdot \beta \simeq 2.5 \times 10^{-16}$~cm, results in $\sigma _\gamma = 2.5 - 0.25 \times 10^{-15}$~cm$^2$, which is in the expected range.
% With typical radiation-induced introduction rates of $(1 - 10)$~cm$^{-1}$ as obtained by microscopic techniques in Ref.~\cite{Davies:2006, Moll:2002}, results in photon cross sections $\sigma _\gamma = (10 - 1) \times 10^{-16}$~cm$^2$, which is in the expected range.
% It is also noted that as measured  $\sigma _\gamma \cdot \beta _{irr}$ decreases with increasing $\lambda$.
%  This is expected, as with decreasing photon energy the fraction of the silicon band-gap which can be reached for electrons in the valence band decreases.
% The same argument is also valid for holes.

 \section{Comparison to edge-TCT results}
  \label{sect:EdgeTCT}

 In Ref.~\cite{Scharf:2018, Feindt:2017} the edge-TCT (Transient-Current-Technique) is used to measure $\lambda _{abs} (\Phi _{eq})$ for proton fluences, $\Phi _{eq}$, between 0 and $13 \times10^{15}~$cm$^{-2}$.
 For edge-TCT~\cite{Kramberger:2010}, focused light from a sub-nanosecond laser is injected through a polished edge into a strip sensor parallel to the surface and perpendicular to the strips, and the current transients are recorded.
 By moving the beam normal to the sensor surface,  the charge-collection efficiency and the electric field as a function of depth, $x$, can be determined~\cite{Kramberger:2014}.
 AC-coupled strip detectors with an active thickness of $285~\upmu$m and a strip pitch of $80~\upmu$m built on $\approx 5~$k$\Omega \cdot$cm $n$-doped silicon were investigated.
 The fourth strip, located $1350~\upmu$m from the sensor edge, and the ninth strip, located $L = 400~\upmu$m from the first strip, were read out.
 The absorption length is obtained from the integral of the current of the first strip, $Q_1$, and of the second readout strip, $Q_2$, using the relation $\lambda _{abs} =  L/\ln(Q_1 / Q_2)$.

 The wavelength of the laser light was $\lambda = 1.052~\upmu$m.
 Because of high leakage currents, the measurements were performed at $- 30~^\circ$C.
 The irradiated sensors were biased with forward voltages, $V_{forw}$, as well as with reverse voltages, $V_{rev}$, between 0 and 1000~V.
 A constant charge ratio, $Q_1(x)/Q_2(x)$, as a function of $x$ was expected.
% As a function of $x$ a constant ratio $Q_1(x)/Q_2(x)$ was expected.
 However, because of a number of experimental problems, this was not the case~\cite{Scharf:2018, Feindt:2017}.
% and therefore only qualitative  conclusions can be made.
 The analysis is performed in the $x$-region, where $Q_1 / Q_2$ is approximately constant.

 It is found that $\alpha _{irr} \propto \Phi _{eq}$, as was already observed for the spectrophotometer measurements.
% For $V_{forw} = 500 - 1000$~V the value of $\alpha _{irr} / (\alpha _{0} \cdot \Phi _{eq}) = (4.2 \pm 1.0) \times 10^{-17}$~cm$^2$,  and for $V_{rev} = 200 - 1000$~V $(5.3 \pm 2.0) \times 10^{-17}$~cm$^2$.
% For $V_{forw} = 500 - 1000$~V the value of $\alpha _{irr} \cdot \lambda _{abs} (0) / \Phi _{eq} = (4.2 \pm 1.0) \times 10^{-17}$~cm$^2$,  and for $V_{rev} = 200 - 1000$~V it is $(5.3 \pm 2.0) \times 10^{-17}$~cm$^2$.
% The large uncertainties reflect the observed increase of $\alpha _{irr} $ with voltage.
% Using the formula $\alpha _0(T) = (T[K]/146.2)^{3.86}$ from Ref.~\cite{Scharf:2018}, one obtains $\alpha _0(T) = 7.1$~cm$^{-1}$ for $\lambda = 1.052~\upmu$m  at $T = 243$~K, and the values $\sigma _\gamma \cdot \beta _{irr} = \alpha _{irr} / \Phi _{eq} = (3.0 \pm 0.7) \times 10^{-16}~$cm for $V_{forw}$, and $(3.8 \pm 1.5) \times 10^{-16}~$cm for $V_{rev}$.
% Using $1/\lambda _{abs} = 6.44$~cm$^{-1}$ for $\lambda = 1.052~\upmu$m at $- 30~^\circ$C for non-irradiated silicon from~\cite{Green:2008}, $\alpha _{irr}/\Phi _{eq} = (2.7 \pm 0.6) \times 10^{-16}~$cm for $V_{forw}$, and $(3.4 \pm 1.3) \times 10^{-16}~$cm for $V_{rev}$.
 The results for the sensor irradiated to $\Phi _{eq} = 9.4 \times 10^{15}$~cm$^{-2}$ are:
 For $V_{forw}$ between 500  and 1000~V the value found for $\alpha _{irr}/\Phi _{eq}$ is $(2.7 \pm 0.6) \times 10^{-16}~$cm, and for $V_{rev}$ between 500  and 1000~V, $(3.4 \pm 1.3) \times 10^{-16}~$cm.
 The large uncertainties reflect the observed increase of $\alpha _{irr} $ with voltage.
 As shown in Fig.~\ref{fig:SigmaBeta}, the results are compatible with the value $(2.9 \pm 0.4) \times 10^{-16}~$cm determined from the spectrophotometer measurements at $+ 20~^\circ$C.
 These results confirm the reduction of $\lambda _{abs}$ due to radiation damage.

  \section{Discussion of the results}
  \label{sect:Discussion}

 The attenuation of radiation in matter as a function of the path-length $x$ follows an exponential law $e^{- \alpha \cdot x} $, with $\alpha = \sum _i (N_i \cdot \sigma _i)$, where $N_i$ is the density of scattering centers and $\sigma _i$ the corresponding cross-sections.
 Analogously one can define
 \begin{equation}\label{equ:IntSig}
  \alpha _{irr} (\Phi _{eq} ) = \sum _i \big( N_i(\Phi _{eq}) \cdot \sigma _i ( \lambda)\big) =
  \Phi _{eq} \cdot \sum _i \big( \beta _i \cdot \sigma _i (\lambda) \big).
 \end{equation}
 The right-hand side of the equation uses the experimental observation of Sect.~\ref{sect:Results} that $\alpha _{irr} \propto \Phi _{eq}$, which allows introducing the introduction rates $\beta _i = N _i/\Phi _{eq}$ for the individual radiation-produced states.

 Typical total radiation-induced introduction rates from microscopic measurements are $(1 - 10)$~cm$^{-1}$ (Ref.~\cite{Moll:2002, Davies:2006}).
 Using the value $\alpha _{irr}/\Phi _{eq} = 3 \times 10^{-16}$~cm from Fig.~\ref{fig:SigmaBeta} gives a range for the average $\sigma $ of $ (3 - 0.3)\times 10^{-16}$~cm$^2$, which is similar to the electron/hole cross-sections obtained from the microscopic measurements.
 It is also noted that the measured $ \sum _i \big( \beta _i \cdot \sigma _i (\lambda) \big) $  decreases with increasing $\lambda$.
 The inter\-pretation is, that due to energy conservation, the fraction of the band-gap which can be reached by electrons from the valence band increases with photon energy.
 A similar argument holds for holes.

 The light absorption for photons with energies close to the band-gap energy, $E_{gap}$, is sensitive to the value of $E_{gap}$.
 Whereas the dependence of $E_{gap}$ on the dopant density, $N_d$, in silicon has been studied in detail~\cite{Aw:1991, Altermatt:2003, Yan:2014}, there are hardly any results on its dependence after irradiation.
 $E_{gap}$ is observed to decrease with $N_d$, however the change is less than $\approx 5$~meV for $N_d < 10^{17}$~cm$^{-3}$.
According to Refs.~\cite{Elliott:1957, Macfarlane:1958, Yu:2010} the dependence of the absorption coefficient for a semiconductor with an indirect band gap can be approximated by $\alpha \propto (\,E_\gamma - E_{gap} + E_{ex} \pm \hbar \omega_p)\,^{2}$ for photon energies $E_\gamma > E_{gap}+\hbar \omega_p$.
The phonon energy required for energy and momentum conservation is $\hbar\omega_p$.
The $+$ sign refers to phonon absorption and the $-$ sign to phonon emission. 
$E_{ex}$ is the exciton binding energy which describes the interaction of the generated electron-hole pair.
Asuming $E_{ex}$ and $\hbar\omega_p$ do not change with irradiation the intercept of a straight-line fit to the linear part of $\sqrt{1/\lambda _{abs}} (E _\gamma)$ with $\sqrt{1/\lambda _{abs}} (E _\gamma = E_g)= 0$ should give $E_g - E_{gap}=const$ similar to Ref.~\cite{Aw:1991}. 
Methods to determine $E_{gap}$ directly are described in Refs.~\cite{Macfarlane:1958, Bludau:1974} but were not applicabale to our data due to the rather large step size of $\Delta \lambda = 1$~nm. 
%However, the determined change $\Delta E_{g}$ should be approximately equal to $\Delta E_{gap}$.
% In Ref.~\cite{Aw:1991} the following  method for the change of $E_{gap}$ with $N_d$ is proposed:
%Plot $\sqrt{1/\lambda _{abs}}$ as a function of $E _\gamma$ and the intercept of the straight-line fit to the linear part with $\sqrt{1/\lambda _{abs}} = 0 $ gives
 %\begin{equation}\label{equ:Eg}
 %  E_g = E_{gap} + \hbar \omega _p,
% \end{equation}
 %with the phonon energy $\hbar \omega _p$.
 %The differences of $E_g$ for different $N_d$ values is a sensitive measure of the change of $E_{gap}$ with $N_d$.
% If the dominating phonons involved in the absorption process are the lowest-energy phonons, the value $\hbar \omega _p = 18$~meV for longitudinal acoustic phonons is expected at room temperature~\cite{Aw:1991}.

  \begin{figure}[!ht]
   \centering
    \includegraphics[width=0.6\textwidth]{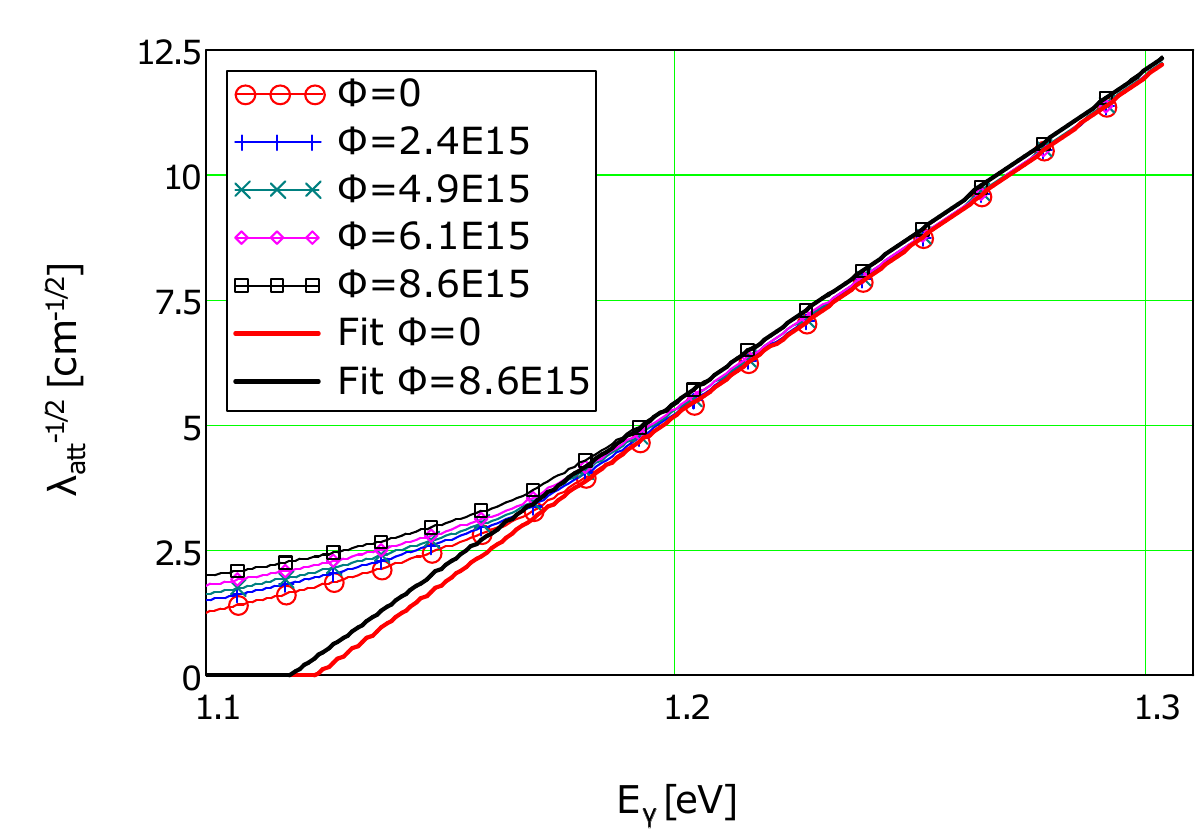}
   \caption{ Plot of $\sqrt{1/\lambda _{abs}}$ as a function of $E _\gamma$ for the different $\Phi _{eq}$-values.
   Straight-line fits in the range $E _\gamma = 1.2 - 1.3$~eV are shown, which allow to determine $E_g$ as a function of $\Phi _{eq}$.  }
  \label{fig:Sqrt}
 \end{figure}

 Fig.~\ref{fig:Sqrt} shows $\sqrt{1/\lambda _{abs}}$ as a function of the photon energy, $E _\gamma$, for the different $\Phi _{eq}$~values.
 For $E_\gamma \gtrsim 1.18$~eV the dependence is linear.
 The intercept of straight-line fits in the range $E _\gamma = 1.2 - 1.3$~eV with  $\sqrt{1/\lambda _{abs}} = 0 $ is $E_g(\Phi _{eq})$.
 For $\Phi _{eq} = 0$ the value $E_g (0) = (1.1230 \times 10^3 ~^{+1.0} _{-0.4} \pm 2.0 \pm 0.5)$~meV is found.
 The first uncertainty is obtained by changing the fit range by $\pm 20$~meV, the second by changing the $\lambda $~scale of the spectrophotometer by $\pm 2$~nm, and the third by calculating the change of $E_{gap}$ for a temperature change by $\pm 2~^\circ $C, using the temperature dependence of $E_{gap}$ of $-0.25$~meV/$^\circ $C at $+20~^\circ $C from Ref.~\cite{Paessler:2002}.
 The systematic effect of a possible local variation of the $\lambda $~scale is not taken into account.
 The statistical uncertainty of the extrapolation is negligible.
 The value of $E_{gap}$ from Ref.~\cite{Paessler:2002} is $E_{gap} = 1.126$~eV with an estimated uncertainty of 1~meV.
 The observed difference is $E_g - E_{gap}$ is $- 3~ ^{+4} _{-5}$~meV.
% compared to $+ 18$~meV expected from longitudinal acoustic phonons.
 %It is assumed that the difference is related to the analysis model used and the neglect of the absorption related to radiation-produced states in the band gap.

  \begin{figure}[!ht]
   \centering
    \includegraphics[width=0.6\textwidth]{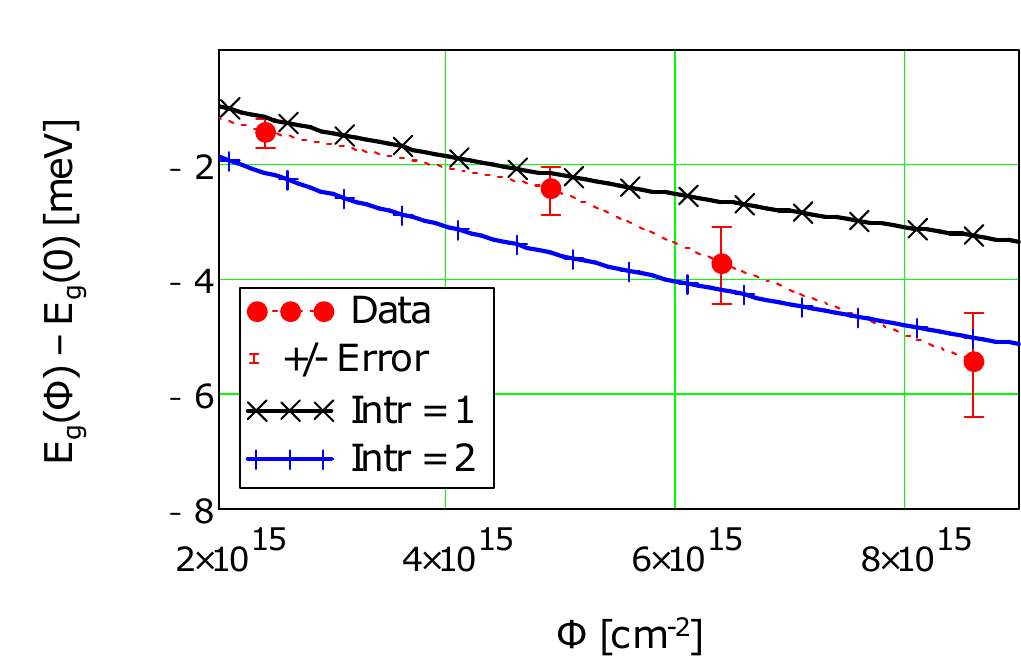}
   \caption{ Difference $E_g(\Phi) - E_g(0) $ obtained from the extrapolation of the straight-line fits shown in Fig.~\ref{fig:Sqrt}.
   As discussed in the text the difference is to a good approximation equal to $\Delta E_{gap}(\Phi _{eq})$, the change of the silicon band-gap with irradiation.
   The two curves are obtained by using Eq.~\ref{equ:YamModel} assuming $N_d = $~\emph{Intr}$ \cdot \Phi _{eq}$ for the two values \emph{Intr} $ = 1$ and $2~$cm$^{-1}$. }
  \label{fig:DEgap}
 \end{figure}

 Fig.~\ref{fig:DEgap} shows the difference $E_g(\Phi _{eq}) - E _g(0) $ as a function of $\Phi _{eq}$.
 The error bars take into account that most of the uncertainties of the $E_g$ determination for the different $\Phi _{eq}$ values are correlated and cancel in the difference.
 Also most of the model uncertainties are expected to cancel in the difference.
 Assuming that $E_{ex}$ and $\hbar \omega _p$ do not change with irradiation, the change of the silicon band-gap with fluence is $\Delta E_{gap}(\Phi _{eq}) = E_g(\Phi _{eq}) - E_g(0)$.
 The data shown in Fig.~\ref{fig:DEgap} suggest that $\Delta E_{gap}(\Phi _{eq}) \propto \Phi _{eq}$.

 In Ref.~\cite{Yan:2014} the following parametrisation for $\Delta E_{gap}$ as a function of doping $N_d$ is given:
 \begin{equation}\label{equ:YamModel}
   \Delta E_{gap}(N_d) = A \cdot \Big( \ln (N_d/N_{ref} ) \Big)^b.
 \end{equation}
 The value of $A$ for $n$-type silicon is $( 3.67 \pm 0.20 ) \times 10^{-5}$~eV if Boltzmann statistics is assumed, and $ ( 4.20 \pm 0.30 ) \times 10^{-5}$~eV for Fermi-Dirac statistics.
 The parameter $N_{ref} = 10^{14}$~cm$^{-3}$, and the exponent $b = 3$.
 To relate the fluence $\Phi _{eq}$ to the number of states, an introduction rate has to be assumed.
 The curves in Fig.~\ref{fig:DEgap} correspond to introduction rates of 1~cm$^{-1}$ and 2~cm$^{-1}$, respectively, assuming the parameters for Boltzmann statistics.
 The values of $\Delta E_{gap}$ for data and model are similar, however the assumption of an introduction rate of 1 to 2~cm$^{-1}$ for damage states which introduce a band gap shift, is completely ad hoc.
 It is also not clear, if Eq.~\ref{equ:YamModel} can be extrapolated to such low concentrations of charged defects.
 In Ref.~\cite{Fan:1956} a linear dependence of $\Delta E_{gap}$ on neutron irradiation fluence with a value
 $\Delta E_{gap} / \Phi  = -2 \times 10 ^{-18}$~eV$\cdot$cm$^2$ is reported.
 For the maximal fluence of the absorption measurement of this paper, $\Phi_{eq} = 8.6 \times 10^{15}$~cm$^{-2}$, the expected value for the band-gap shift would be $\Delta E_{gap} = -17$~meV, significantly larger than what is observed, assuming a hardness factor of 1 for the irradiations of Ref.~\cite{Fan:1956}.
 Additional work is required to clarify the situation.

 It should also be noted that most of the damage by 24~GeV/c protons results in defect clusters, whereas the dopants are atoms located at the lattice points.
 Ref.~\cite{Gossick:1959} proposes that cluster defects can change the potential locally, which results in a local change of the band gap.
 In Ref.~\cite{Donegani:2018} it is shown that the activation energy of states in clusters can be several meV lower than for point defects.
 Such effects can possibly explain the $\Phi _{eq}$-dependence and the magnitude of $\Delta E_{gap}$.

  \section{Conclusions and outlook}
  \label{sect:Conclusions}

 The change of the light-absorption length in silicon, $\lambda _{abs} (\Phi _{eq}, \lambda)$, due to radiation damage by 24~GeV/c protons for 1~MeV neutron-equivalent fluences, $\Phi _{eq}$, up to $8.6\times 10^{15}$~cm$^{-2}$ has been measured using a spectrophotometer.
 The measurements were performed at $+ 20~^\circ$C for wavelengths, $\lambda $, between 0.95 and $1.30~\upmu $m.
 In the $\Phi _{eq}$-range investigated it is found that
 $\alpha _{irr} (\Phi _{eq}, \lambda) = 1/\lambda _{abs} (\Phi _{eq}, \lambda) - 1/\lambda _{abs} (0, \lambda)$ is proportional to $\Phi _{eq}$.
% The quantity $\alpha _{irr}/ \Phi _{eq}$ is interpreted as the product of the photon cross-section and the density of radiation-induced states in the silicon band-gap relevant for a given $\lambda $.
 A phenomenological parametrisation of $\alpha _{irr}(\lambda )/\Phi _{eq}$ is presented, which, together with the $\lambda _{abs} (0) $~data for non-irradiated silicon of Ref.~\cite{Green:1995, Green:2008}, allows to calculate $\lambda _{abs} (\Phi _{eq}, \lambda)$.
 Edge-TCT measurements of strip sensors irradiated to $\Phi _{eq} = 9.4 \times 10^{15}$~cm$^{-2}$ performed at $ - 30~^\circ$C with light of $\lambda = 1.052~\upmu$m confirm the observed decrease of $\lambda _{abs}$ with $\Phi _{eq}$.
% For $\alpha _{irr} / \Phi _{eq}$ values compatible with the spectrophotometer results are found.
 The quantity $\alpha _{irr}/ \Phi _{eq}$ is interpreted as $\sum _i \big( \beta _i \cdot \sigma _i (E _\gamma) \big)$, the sum over the products of the introduction rates of the radiation-induced damage states $\beta _i$ times the corresponding photon cross-sections $\sigma _i (E_\gamma)$.

 From the measured $\lambda _{abs} (\Phi _{eq}, \lambda )$ the change of the band-gap energy, $\Delta E_{gap}(\Phi _{eq})$, as a function of $\Phi _{eq}$ is extracted.
 $\Delta E_{gap}(\Phi _{eq})$ is approximately proportional to $\Phi _{eq}$ and reaches a value of  $\approx - 5$~meV at the highest fluence of $\Phi _{eq} = 8.6 \times 10^{15}$~cm$^{-2}$.

 The measurements and results presented are a step towards a systematic study of the radiation-induced change of light absorption in silicon.
 A next step could be to determine $\alpha _{irr} $ as a function of temperature using spectrophotometer measurements.
 With edge-TCT $\alpha _{irr} $ can be studied as a function of temperature, electric field and current.
 These parameters influence the occupancy of the radiation-induced states and therefore the photon cross-sections $\sigma _i$.

 As the $\sigma _i (E_\gamma )$ times the densities of damage centers, $N_i$, enter into the calculation of the specific energy loss of charged particles, $\mathrm{d}E/\mathrm{d}x$~\cite{Bichsel:1998, Bichsel:2006}, the generation of charge carriers per unit length may be different for non-irradiated and irradiated sensors.
 However, the expected effect is estimated to be negligibly small.
 For the band-gap narrowing due to radiation damage further studies are required to verify the results presented in this paper and refine the analysis methods.
% Finally the results of $\alpha _{irr}$ can be compared to a possible change of the generation of electron-hole pairs by charged particles.
% As the lower limit of the virtual photon cross-section decreases because of the radiation-induced states in the band-gap, an increase of the energy loss of charged particles and of the number of generated electron-hole pairs is expected.

%\paragraph*{Acknowledgement}
%\section{Acknowledgement}
%\section*{Acknowledgement}
% \label{sect:Acknowledgement}
% We would like to thank .

\section*{Acknowledgements}
 \label{sect:Acknowledgement}
  We thank Philip Metz and Alexander Heuer from the group of G\"unter Huber of the Institute of Laser Physics of Hamburg University for their help with the photospectrometer measurements,  Eckhart Fretwurst for providing the silicon samples, and Eckhart Fretwurst, Erika Garutti and Joern Schwandt for stimulating discussions.
  The project was supported by the HGF Alliance \emph{Physics at the Terascale} and  the H2020 project AIDA-2020, GA no. 654168.

%\newpage
 \input{bibliography}

  \label{sect:Bibliography}

%\bibliographystyle{unsrt}
%\addcontentsline{toc}{section}{\refname}

%\bibliography{bib/bib}

%\bibliographystyle{unsrtnat}
\end{document}

%% file: bibliography.tex
%
%   Nuclear Instruments and Methods in Physics Research A 409 (1998) 184 -- 193.

\section{List of References}